\documentclass[reprint,aps,pra,nofootinbib,superscriptaddress]{revtex4-2}
\usepackage{braket}
\usepackage{slashed}
\usepackage{graphicx}
\usepackage{xcolor}
\usepackage{enumitem}
\usepackage{siunitx}
\usepackage{mathtools}
\usepackage{gensymb}
\usepackage{dsfont}
\usepackage{textcomp}
\usepackage{amsmath,amsthm,amsfonts,amscd,amssymb} 
\usepackage{comment}

\newcommand\blfootnote[1]{%
  \begingroup
  \renewcommand\thefootnote{}\footnote{#1}%
  \addtocounter{footnote}{-1}%
  \endgroup
}

\usepackage{tikz} 
\usetikzlibrary{shapes,arrows,positioning,automata,backgrounds,calc,er,patterns}
\usetikzlibrary{decorations,decorations.markings,decorations.pathreplacing,decorations.pathmorphing}
\usepackage{relsize}

\newcommand{\be}{\begin{equation}}
\newcommand{\ee}{\end{equation}} 
\newcommand{\mb}{\mathbf}

\begin{document}

\title{Note on quantum optimal measurements of CMB spectral distortions}

\author{Roshni Singh*}
\affiliation{Physics Division, Lawrence Berkeley National Laboratory, Berkeley, CA 94720, USA}
\affiliation{Department of Physics, University of California, Berkeley, CA 94720, USA}

\author{Jacques Delabrouille}
\affiliation{Physics Division, Lawrence Berkeley National Laboratory, Berkeley, CA 94720, USA}
\affiliation{CNRS-UCB International Research Laboratory, Centre Pierre Binétruy, IRL2007, CPB-IN2P3, Berkeley, CA 94720, USA}

\blfootnote{* roshni$\_$singh@berkeley.edu, carney@lbl.gov}

\author{Daniel Carney*}
\affiliation{Physics Division, Lawrence Berkeley National Laboratory, Berkeley, CA 94720, USA}
\affiliation{Department of Physics, University of California, Berkeley, CA 94720, USA}

\begin{abstract}
As quantum sensing techniques continue to push the frontier of detection of ultra-weak signals, one may wonder if a sophisticated quantum strategy could be used to enable measurement of spectral distortions in the cosmic microwave background. Using quantum Fisher information techniques, we prove that the optimal strategy is simple frequency-binned intensity measurement. This proof strategy also enables us to provide numerical estimates for the information-theoretic minimum observation time required to observe the $\mu$- and $y$-distortions expected from $\Lambda$CDM cosmology. As case studies, we compare these results to the  PIXIE and SPECTER mission concepts, and show that they are within a few orders of magnitude of this fundamental limit.
\end{abstract}
\maketitle

\section{Introduction}
\label{intro}

Quantum measurement techniques are enabling measurements of ultra-weak signals in fundamental physics, most notably including gravitational waves~\cite{ganapathy2023broadband,Beckey:2023shi}. One particularly intriguing set of such weak signals are possible spectral distortions of the cosmic microwave background, i.e., deviations from an exact blackbody spectrum. The concordance $\Lambda$CDM model predicts a small but non-zero level of these distortions, and measuring them in detail would provide a wealth of valuable information about both the early and later universe~\cite{sunyaev1969distortions, fu2021unlocking, Fixsen_2011, mroczkowski2019astrophysics}. The best current observational limits were set in the 90's by COBE/FIRAS~\cite{Fixsen:1996nj}, and a number of upgraded mission concepts have been proposed in recent years~\cite{2014JCAP...02..006A,2021ExA....51.1515C,2021ExA....51.1471D,10.1117/12.2231090,kogut2025primordial,sabyr2025specter}. 

In this note, we study the elementary question of whether there is a fundamentally quantum-optimal measurement strategy for these CMB spectral distortions. The answer turns out to be simple. We prove that frequency-binned photon counting is the optimal measurement scheme for measuring spectral distortions. The proof is based on a result of Nair and Tsang on the optimal strategy for stellar temperature measurements~\cite{nair2015quantum}. We compute the quantum Fisher information for spectral distortions, which provides a bound on the total information that can be extracted from a signal using any possible quantum measurement strategy \cite{fiderer2021general,liu2020quantum}, and show that it is saturated by binned photon counting. This shows, for example, that using entanglement-based measurements or other sophisticated quantum schemes cannot improve the sensitivity over photon counting. 

\begin{figure}[t]
    \centering
    \includegraphics[width=0.9\linewidth]{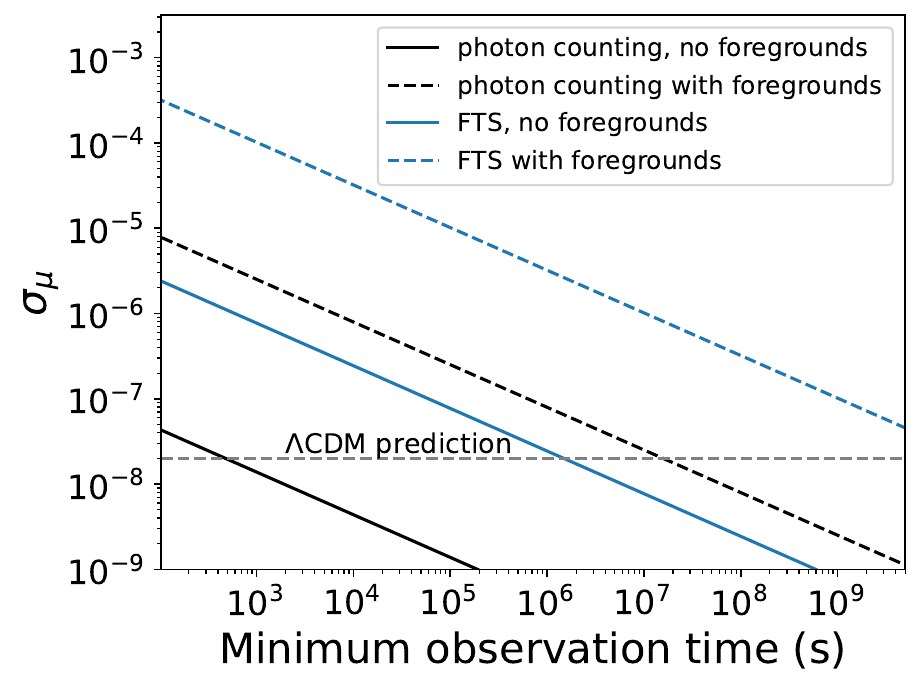}
    \caption{Obtainable measurement uncertainty $\sigma_{\mu}$ in the $\mu$-distortion as a function of observation time. The various lines represent idealized photon-counting and Fourier Transform Spectrometer (FTS) detectors, using fiducial instrumental parameters listed in Table ~\ref{tab:param_values} and foreground parameters in Table~\ref{tab:foregrounds}.}
    \label{fig:missions_compare}
\end{figure}

Quantitatively, this calculation also provides a universal benchmark against which any spectral distortion measurement can be compared. As a concrete application, we estimate the minimum observation time required to measure $\mu$-distortions at the $\mu \approx 2 \times 10^{-8}$ level predicted by $\Lambda$CDM cosmology~\cite{chluba2016spectral}, as shown in Fig.~\ref{fig:missions_compare}. Using ideal binned photocounting in a satellite mission with similar geometric parameters as the PIXIE or SPECTER concepts, we find that the minimum observation time is on the order of months when foregrounds are considered. Foregrounds and detector parameters strongly affect this forecast; we provide simple code at~\cite{spectral_distortions_estimates_roshni} which can be used to vary our specific choices.

We end with a few comments comparing this idealized detection scheme to proposed mission concepts ~\cite{10.1117/12.2231090,kogut2025primordial,sabyr2025specter}. We find that binned photon counting, although possibly harder to implement in practice, is in principle superior to Fourier transform spectroscopy (FTS) techniques, since the latter spreads the photon shot noise across all bins while the former does not. These conclusions provide further motivation for the development of absolutely calibrated frequency-resolved intensity detectors. 

\section {Optimality of photon counting}
\label{optimality_section}
In this section, we show that frequency-binned photon counting (intensity measurement) is the optimal measurement strategy. To do this, we use a quantity called the quantum Fisher information. Classical Fisher information determines the accuracy with which we can estimate a parameter from data taken using a specified measurement protocol. In contrast, the quantum Fisher information determines the accuracy with which we can estimate a parameter from data, optimized over \emph{all possible} quantum measurement protocols. 

The proof strategy is straightforward, and closely follows a related problem in astronomy studied by Nair and Tsang~\cite{nair2015quantum}. We calculate the quantum Fisher information for the problem of estimating the spectral distortion parameters, the classical information for the same problem assuming one chooses photon counting measurements, and then show that this classical Fisher information is equal to the quantum Fisher information. This establishes that photon counting is the optimal measurement strategy to provide a bound on CMB spectral distortions. This answer is not surprising: the basic task is to check if a measured intensity curve follows a precisely thermal spectrum. Our arguments here rigorously obviate the need for any more sophisticated measurement strategies.

Before we go through the proof, let us be clear about the assumptions that go into this idealized calculation. We assume ideal detectors, which for photon counting means we assume perfect optical efficiency (all incoming photons are detected) and ignore dark counts. This is the true quantum limit for photodetection, because an intensity measurement does not suffer from vacuum quantum fluctuations in the detector. This is in contrast to, for example, homodyne measurements, which have a ``half-photon'' of noise even in the vacuum~\cite{PhysRevD.23.1693} (although this can in principle be alleviated by methods like squeezing~\cite{PhysRevD.23.1693,PhysRevLett.55.2409}). Our goal here is simply to provide a fundamental benchmark to compare realistic experiments against, but we make a few comments on dark counts at the end of the paper. 

We explain our model for the CMB photon signal in Sec.~\ref{modeling_section} below. We then give the proof of optimality both without foregrounds as a warmup in Sec.~\ref{classical_fisher_section}, and with foregrounds in Sec.~\ref{foregrounds_section}. In Sec.~\ref{numerics} we then use this proof strategy to compute what is called the Cram\'er-Rao bound for the problem, which allows us to make numerical estimates for the observation time required to reach a given precision on the spectral distortion parameters. Here we focus mostly in the case of $\mu$-distortions since it is the most difficult experimental target, but the formalism also works identically for the $y$-distortions.

\subsection{Modeling  CMB photons}
\label{modeling_section}
In order to compute the quantum Fisher information, we need to represent the quantum state of the CMB radiation as a density matrix. Let $\mb{k}$ denote a Fourier mode of the radiation field. These are uncorrelated, so the total density matrix is just a product over all modes,
\begin{equation}
\label{rho_tot}
\rho (\boldsymbol{\theta}) = \bigotimes_{\mb{k}} \rho_{\rm{th,\nu_\mb{k}}}( \boldsymbol{\theta}).
\end{equation}
Here the free parameters we are attempting to constrain are listed as a vector $\boldsymbol{\theta} = [T,\mu,y,\theta_{e}]$, where $T$ is the temperature of the CMB, and $\mu$, $y$ and $\theta_e$ characterize spectral distortions. With the simplifying assumption that the $\mu$-distortions are isotropic and frequency independent, $\mu$ acts as a chemical potential and shifts the expected number of photons at each frequency~\cite{sunyaev1969distortions}.

We model the density matrix of each mode of light, with frequency $\nu_{\mb{k}} = |\mb{k}|$, as the usual thermal spectrum including a chemical potential and the $y-$distortion signal:
\begin{equation}
\label{single_mode_density_matrix} 
 \rho_{\rm th,\nu_\mb{k}}(\boldsymbol{\theta}) = \frac{1}{\bar n_\mb{k}(\boldsymbol{\theta})+1}\sum_{n_\mb{k}=0}^{\infty}\left(\frac{\bar n_\mb{k}(\boldsymbol{\theta})}{\bar n_\mb{k}(\boldsymbol{\theta})+1}\right)^{n_\mb{k}}|n_\mb{k}\rangle\langle n_\mb{k}|.
\end{equation}
Here $\bar n_\mb{k}(\boldsymbol{\theta})$ is the mean number of photons at frequency $\nu_\mb{k}$ at a given temperature, chemical potential and  $y-$distortion
\begin{gather}
\label{bar_n}
    \bar n_\mb{k}(\boldsymbol{\theta}) = \frac{1}{e^{\frac{\nu_\mb{k}}{T}+\mu}-1} +\Delta n_y\bigg(x=\frac{\nu_\mb{k}}{T},y,\theta_e\bigg),
\end{gather}
and $\ket{n_{\mb{k}}}$ is the Fock state associated with $n_{\mb{k}}$ photons in this particular mode. $\Delta n_y(x,y,\theta_e)$ accounts for the contribution of the $y-$distortion to the mean number of photons (for details, see appendix ~\ref{y-distortions}). 
\begin{figure}[t]
    \centering
    \includegraphics[width=\linewidth]{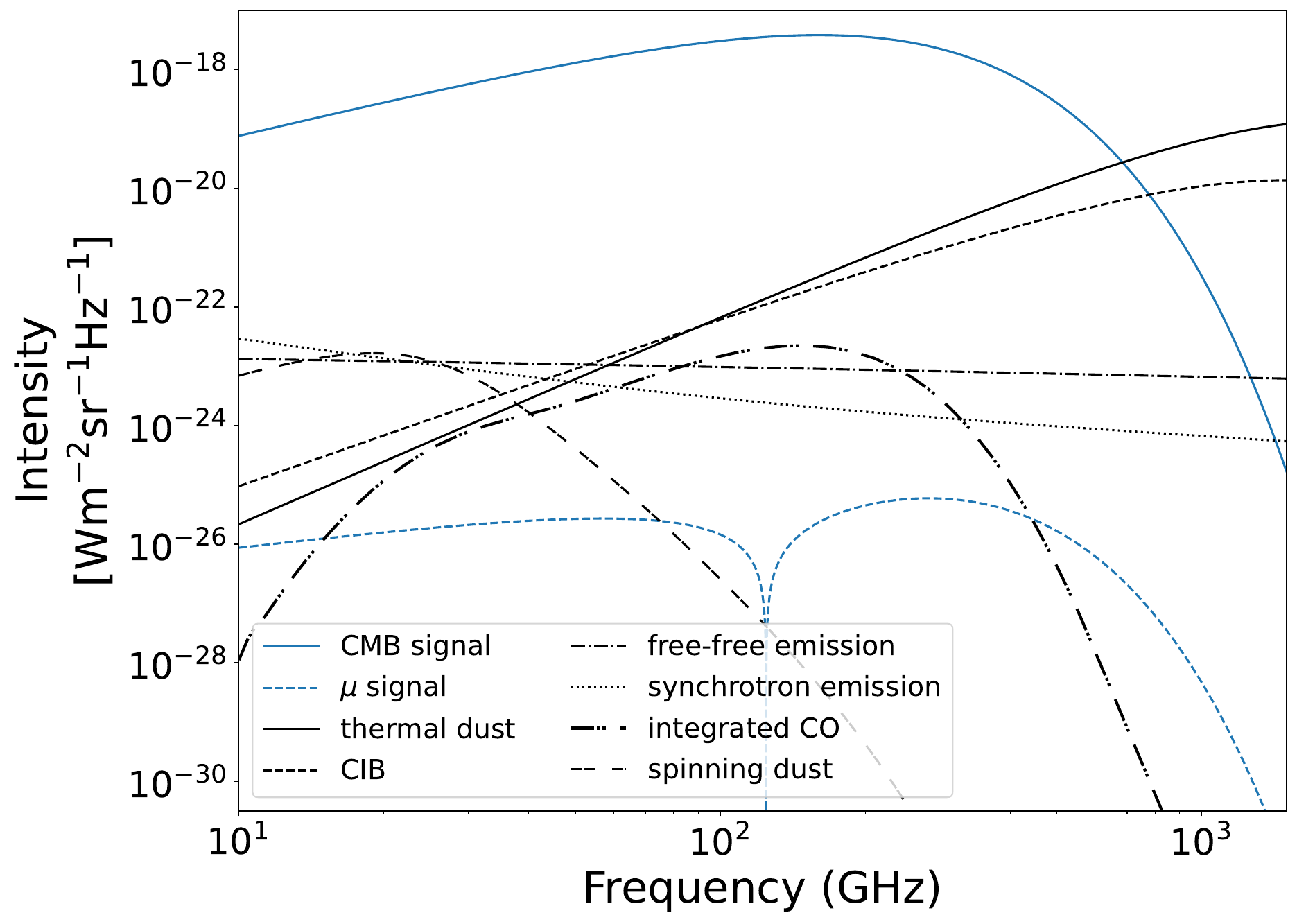}
    \caption{Fiducial models of the spectra of the CMB radiation and foregrounds used in this paper. The foreground models are taken from~\cite{abitbol2017prospects}. The $\mu$ signal line is the difference $\Delta I = |I_{\mu=0} - I_{\mu \neq 0}|$ between spectra with $\mu=0$ and non-zero $\mu$, shown here with $\mu=2\times10^{-8}$.}
    \label{fig:CMB_signal}
\end{figure}

We will assume that our detector collects and independently measures the radiation in multiple frequency bands with central frequencies $\nu_1,\nu_2,...\nu_N$ and bandwidth $\Delta \nu$. We will show below that this is sufficient for an optimal measurement because the Fourier modes of the signal are uncorrelated over frequencies, Eq.~\eqref{rho_tot}.

When considering the input radiation field incident on a finite sized detector we need to account for multiple spatiotemporal modes in each frequency band  ~\cite{dandliker1999concept,sova2025etendue,fabre2020modes}. To count the number of transverse spatial modes, we take the ratio of the total area of the detector, $A$, to the coherence area of a single mode, $\lambda_{\mb{k}}^2\Omega^{-1}$~\cite{Mandel_Wolf_1995_ch4,zmuidzinas2003thermal}. Modes outside this coherence area are uncorrelated with those outside. Here $\lambda_{\mb{k}}$ is the wavelength of the radiation and $\Omega$ is the solid angle subtended by the detector. The number of temporal modes is the ratio of the total observation time to the coherence time, $\tau_c \approx (\Delta\nu)^{-1}$~\cite{nair2015quantum,Born_Wolf_2019}.
The total number of modes in any frequency band accounting for both polarization modes is:
\begin{align}
\label{mode_count}
    N (\nu_{\mb{k}})= \frac{2 A \Omega \nu_{\mb{k}}^2 \Delta\nu}{c^2} \times T_{obs}.
\end{align}
Given these definitions, we can decompose the quantum state of the radiation field on the detector into a product over these measured bands accounting for each independent spatiotemporal mode,
\begin{equation}
\label{rho_modes}
\rho (\boldsymbol{\theta}) = \bigotimes_{\mb{k}, m\in \mathcal{M_{\mb{k}}}}\rho^{(m)}_{\rm{th,\nu_\mb{k}}}( \boldsymbol{\theta}) \bigotimes \rho_{\rm unmeasured},
\end{equation}
where $\mathcal{M_\mb{k}}$ denotes the full set of spatiotemporal modes associated with a Fourier mode with frequency $\nu_{\mb{k}}$. 
We assume that within each band, the mean number of photons is roughly constant, so that the mean number of photons in each mode in the input field for the band can be represented by Eq.~\eqref{bar_n}. 
Since this holds true for every mode $ m \in \mathcal{M_{\mb{k}}}$, we henceforth drop the label $m$ for the spatiotemporal modes.
The total Fisher information over the radiation spectrum is obtained by summing the quantum Fisher information over all the spatiotemporal modes~\cite{fabre2020modes, nair2015quantum} for each band.  The unmeasured part of the input radiation does not contribute to the Fisher information.
\subsection{Quantum Fisher information of CMB photons}
\label{quantum_fisher}
For a multiparameter density matrix $\rho(\boldsymbol{\theta})$, the quantum Fisher information matrix is ~\cite{liu2020quantum}
\begin{equation}
\label{def_fisher_eigenspectrum}
    F_Q^{ij}(\boldsymbol{\theta}) = 2\sum_{m,l\\\lambda_m+\lambda_l\neq 0} \frac{Re (\bra{m} \partial_{\theta_i} \rho(\boldsymbol{\theta})  \ket{l}\bra{l}\partial_{\theta_j} \rho(\boldsymbol{\theta}) \ket{m})}{\lambda_m + \lambda_l}.
\end{equation}
Here $\lambda_m,\lambda_l$, $\ket{m},\ket{l}$ are the eigenvalues and eigenvectors of the density matrix.
We can compute the total Fisher information matrix by summing over the Fisher information of each independent band $F_{Q,\nu_\mb{k}}^{ij}(\boldsymbol{\theta})$, weighted by the number of modes in that band as given by Eq. ~\eqref{mode_count}:
 \begin{equation}
 \label{fisher_tot}
     F_Q^{ij}(\boldsymbol{\theta}) = \sum_{\nu_\mb{k}\in \boldsymbol{\nu}}N({\nu_\mb{k}}) F_{Q,\nu_\mb{k}}^{ij}(\boldsymbol{\theta}).
 \end{equation}
To compute $F_{Q,\nu_k}^{ij}(\boldsymbol{\theta})$ , 
we note that the single mode density matrix $\rho_{\rm{th},\nu_\mb{k}}(\boldsymbol{\theta})$ is diagonal in the Fock basis. Since the dependence of $\rho_{\rm{th,\nu_{\mb{k}}}}$ on $\boldsymbol{\theta}=[T,\mu,y,\theta_e]$ is only through its dependence on $\bar{n}_{\mb{k}}(\boldsymbol{\theta})$, the derivatives of the density matrix with respect to the parameters $\theta_i \in \boldsymbol{\theta}$, are also diagonal in the Fock basis. 
This enables us to easily compute the Fisher information matrix elements for a given mode:
\begin{align}
\label{quantum_fisher_mode_result}
 F_{Q,\nu_{\mb{k}}}^{ij} (\boldsymbol{\theta})
= \frac{1}{\bar{n}_{\mb{k}}(\boldsymbol{\theta})(\bar{n}_{\mb{k}}(\boldsymbol{\theta})+1)}\left(\frac{\partial\bar{n}_{\mb{k}}(\boldsymbol{\theta})}{\partial \theta_i}\right)\left(\frac{\partial \bar{n}_{\mb{k}}(\boldsymbol{\theta})}{\partial{\theta_j}}\right).
\end{align}
To get this, we essentially just used the chain rule in the form $\partial \rho/\partial \theta = (\partial \rho/\partial n) (\partial n/\partial \theta)$.

Eq.~\eqref{quantum_fisher_mode_result} is the central result of this section. We could evaluate this formula more explicitly, but for the purposes of proving that photon counting is optimal, we will see that this form is sufficient. 

\subsection{Classical Fisher information for photon counting measurements }
\label{classical_fisher_section}
We now calculate the classical Fisher information associated with photon counting measurements, and show that this saturates the quantum Fisher information. Since the quantum Fisher information is the maximum attainable information over all measurement strategies, this proves the optimality of photon counting to measure the spectral distortions. 

The multiparameter classical Fisher information matrix for a random variable, $X$, with a probability distribution $p(X|\boldsymbol{\theta})$ which depends on the estimator $\boldsymbol{\theta}$ is defined as:
\begin{align}
\label{classical_fisher_def}
 F_{cl}^{ij} (\boldsymbol{\theta}) = \sum_{X} \frac{1}{p(X|\boldsymbol{\theta})}\left(\frac{\partial}{\partial \theta_i}p(X|\boldsymbol{\theta})\right)\left(\frac{\partial}{\partial \theta_j}p(X|\boldsymbol{\theta})\right).
 \end{align}
 Just like the quantum Fisher case, the total classical Fisher information across all frequency bands is just a sum:
\begin{equation}
\label{tot_classical_fisher}
    F_{\rm{cl}}^{ij}(\boldsymbol{\theta}) = \sum_{\nu_\mb{k}\in \boldsymbol{\nu}} N(\nu_\mb{k}) F_{\rm{cl,\nu_{\mb{k}}}}^{ij}(\boldsymbol{\theta}).
\end{equation}
For an experiment performing photon counting measurements, the random variable is the number of photons in a particular mode, $n_{\mb{k}}$.
The classical probability distribution for getting $n_\mb{k}$ photons from the CMB radiation given a certain value of $\boldsymbol{\theta}$ is
\begin{equation}
\label{p_m_given_n}
    p(n_{\mb{k}}|\boldsymbol{\theta})= \frac{1}{\bar{n}_{\mb{k}}(\boldsymbol{\theta})+1}\left(\frac{\bar{n}_{\mb{k}}(\boldsymbol{\theta})}{\bar{n}_{\mb{k}}(\boldsymbol{\theta})+1}\right)^{n_{\mb{k}}}.
\end{equation}
The classical Fisher information with respect to $\boldsymbol{\theta}$ can then be computed to be:
\begin{align}
\label{single_mode_classical_fisher}
    F_{\rm{cl},\nu_{\mb{k}}}^{ij} (\boldsymbol{\theta}) = \frac{1}{\bar{n}_{\mb{k}}(\boldsymbol{\theta})(\bar{n}_{\mb{k}}(\boldsymbol{\theta})+1)}\left(\frac{\partial \bar{n}_{\mb{k}}}{\partial \theta_i}\right)\left(\frac{\partial \bar{n}_{\mb{k}}}{\partial \theta_j}\right).
\end{align}
Comparing with the quantum Fisher result, Eq.~\eqref{quantum_fisher_mode_result}, we see that the classical Fisher information associated with photon counting is precisely the quantum Fisher information. Since the quantum Fisher information is already optimized over all possible measurement strategies, we conclude that photon counting is the optimal strategy for measuring $\mu$-distortions.

In hindsight, this is natural since the density matrix is diagonal in the photon number Fock basis, and depends on all other parameters through their dependence on $\bar{n}_{\mb{k}}$. The density matrix has no phase coherence in the Fock basis, so phase sensitive measurements yield no additional information. Essentially, for a density matrix diagonal in Fock space, the optimal measurement strategy is photon counting, and information about the values of $\theta_i \in \boldsymbol{\theta}$ is simply classical post-processing. 
\subsection{Foreground Inclusion}
\label{foregrounds_section}
In addition to the photons from the CMB, foregrounds such as emission from thermal dust, synchrotron radiation, free-free emission, spinning dust, the cosmic infrared background etc. also contribute to the photons collected at the detector. Here we show that  photon counting remains the optimal strategy for detecting $\mu$-distortions even in the presence of these foregrounds. On the other hand, including foregrounds decreases the Fisher information per frequency mode, which will affect the numerical estimates for observation time given in Sec.~\ref{numerics}.

The proof of optimality follows the same basic strategy as above. We compute the quantum Fisher information with the foregrounds included, the classical Fisher information with the foregrounds and assuming photon counting measurements, show that they are equal.

Once we include foregrounds, the probability to detect a total of $n_{\mb{k}}$ photons at frequency $\nu_\mb{k}$ is \cite{Mandel_Wolf_1995_ch11,Mandel_Wolf_1995_ch13}

\begin{gather}
\label{p_tot_main}
    p_{\rm{tot}}(n_\mb{k}|\boldsymbol{\theta}) =  \frac{1}{\bar{n}_{\rm{tot},\mb{k}}(\boldsymbol{\theta})+1}\left(\frac{\bar{n}_{\rm{tot},\mb{k}}(\boldsymbol{\theta})}{\bar{n}_{\rm{tot},\mb{k}}(\boldsymbol{\theta})+1}\right)^{n_{\mb{k}}}, \hspace{2 pt}\rm{with}
    \\ 
    \bar{n}_{\rm{tot},\mb{k}} (\boldsymbol{\theta})= \bar{n}_{s,\mb{k}}(\boldsymbol{\theta})+\bar{n}_{f,\mb{k}}(\boldsymbol{\theta}). \nonumber
\end{gather}

where $\bar{n}_{s,\textbf{k}}(\boldsymbol{\theta})$ is the average number of signal CMB photons and $\bar{n}_{f,\mb{k}}(\boldsymbol{\theta})$ is the average number of foreground photons at frequency $\nu_{\mb{k}}$. Correspondingly, the density matrix for the radiation incident on the detector takes the simple diagonal form:
\begin{align}
\label{rho_foregrounds}
    \rho_{\rm{tot}}(\boldsymbol{\theta}) \hspace{-2 pt}= \hspace{-10.5 pt}\bigotimes_{\mb{k},m\in \mathcal{M}_\mb{k}}\hspace{-6.5 pt}\left(\sum_{n_{\mb{k}}=0}^{\infty} \hspace{-1 pt}p_{\rm{tot}}(n_{\mb{k}},\boldsymbol{\theta}) \ket{n_{\mb{k}}}\bra{n_{\mb{k}}}\right)\hspace{-2 pt}\bigotimes \rho_{\rm{unmeasured}}.
\end{align}
Here, the Fisher parameter list $\boldsymbol{\theta} = [T,\mu,y,\theta_e,\ldots]$ now contains all the foreground parameters as well as the signal ones. However, since the density matrix is still diagonal in the photon number basis, the proof that photon counting is optimal follows exactly the same logic as the foreground-free case of the previous section. 
\begin{table}[t!]
    \centering 
    \begin{tabular}{|c|c|c|}
    \hline    
   \textbf{Parameter} & \textbf{Value} &  \textbf{Source}\\
   \hline
$\mu$ & $2\times 10^{-8}$ &  Value from $\Lambda$CDM~\cite{kogut2025primordial}\\
    \hline
    $y$ & $1.77 \times 10^{-6}$ & Kogut et al.~\cite{kogut2025primordial} \\ 
    \hline
    $\theta_e$ & $2.5\times 10^{-3}$  & Abitbol et al.~\cite{abitbol2017prospects}
    \\
    \hline
    $\omega_2^{\rm{eSZ}} $ & 1.152 & Abitbol et al.~\cite{abitbol2017prospects}\\
    \hline
    Frequency & 10 GHz \rm{to} 2 THz & spans CMB peak,  \\ 
    range & & foregrounds\\
    \hline
    $\Delta \nu$ & 1 GHz & High frequency\\ & & resolution\\
    \hline
     \'Etendue $(A\Omega)$ & 20 $\rm{cm}^2$ $\rm{sr}$ &  order of magnitude  \\
      (per horn)& & estimate\\
    \hline
    Number & 10 & order of magnitude \\
 of horns  & & estimate\\
    \hline
    \end{tabular}
    \caption{Values of fiducial signal and detector parameters used in Fig.~\ref{fig:missions_compare}.}
\label{tab:param_values}
\end{table} 
\subsection{Numerical estimates of observation time}
\label{numerics}
In addition to simply proving that photon counting is the optimal measurement strategy, the Fisher information calculation can be used to estimate the information-theoretic limit on the minimum observation time to search for spectral distortions at some desired sensitivity $\sigma_{\mu}$. This follows from the quantum Cram\'er-Rao bound, which states that the minimum possible error on inference of an unknown parameter is given by $\Delta \theta^2 \gtrsim 1/N F$, where $N$ is the number of measurements and $F$ is the quantum Fisher information.

More precisely, given the quantum Fisher information, the Cram\'er-Rao bound  provides a lower bound on the covariance matrix, $[\mathrm{cov}(\boldsymbol{\theta},\{\Pi_y\})]$, of unbiased estimators of the density matrix parameters ~\cite{liu2020quantum}:
\begin{equation}
\label{cramer_rao_bound}
    \mathrm{cov}(\boldsymbol{\theta}, \{\Pi_y\}) \geq \frac{1}{N} 
    \mathcal{I}^{-1}(\boldsymbol{\theta},\{\Pi_y\}) \geq \frac{1}{N} {F_Q}^{-1}(\boldsymbol{\theta}).
\end{equation}
Here $ \{\Pi_y\}$ represents a specific measurement protocol. Once a measurement protocol such as photon counting or homodyne measurement is fixed, the measurement outcomes associated with it are represented by the set of matrices $\{\Pi_y\}$. 
$\mathcal{I}(\boldsymbol{\theta},\{\Pi_y\})$ is the classical Fisher information matrix of the multiparameter probability distribution of the outcomes when considering a specific measurement protocol represented by $\{\Pi_y\}$. $N$ is the number of independent repetitions of the experiment. In our context, each spatiotemporal mode in a frequency band is an independent repetition of the experiment. Therefore, $N$ is implicitly taken into account when calculating the total Fisher information, Eq.~\eqref{fisher_tot}, which scales linearly with the observation time.

The variance of any parameter $\theta_i\in \boldsymbol{\theta}$ is lower bounded by the corresponding diagonal entry in the covariance matrix. From Eq. ~\eqref{cramer_rao_bound}, we see that the covariance matrix, and hence the variance of any parameter that the density matrix depends on varies inversely with the total quantum Fisher information, $F_{Q}(\boldsymbol \theta)$.  Using Eqs. ~\eqref{mode_count} and ~\eqref{fisher_tot}, we can rewrite the Cram\'er-Rao bound to find the minimum observation time for a desired value of variance $\sigma^2_\mu$ as
\begin{equation}
\label{T_obs_ineq}
    T_{obs} \geq \frac{1}{\sigma_{\mu}^2}\left(\frac{2A\Omega\Delta\nu}{c^2} \right)^{-1}\left(\sum_{\nu_k\in \boldsymbol{\nu}} \nu_k^2F_{Q,\nu_k}\right)^{-1}_{\mu \mu}.
\end{equation}
This result and Eq.~\eqref{quantum_fisher_mode_result} provide one of our core results: the information-theoretic limit on observation time for a given level of spectral distortions.

In Fig.~\ref{fig:missions_compare}, we use Eqs.~\eqref{quantum_fisher_mode_result} and \eqref{T_obs_ineq} to numerically estimate the time required to observe $\mu$ to a desired level of precision $\sigma_{\mu}$ with some fiducial detector and foreground parameters listed in Tables~\ref{tab:param_values} and~\ref{tab:foregrounds}. The FTS sensitivities are calculated using the method in Appendix ~\ref{fts_section}. The required observation time is sensitive to both the instrumental parameters and foreground models. To facilitate experimentation with these parameters based on current and future mission proposals, we provide the code to produce numerical observation times on github \cite{spectral_distortions_estimates_roshni}. 
\section{Outlook and conclusions}
Our central result is that frequency-binned photon counting is the fundamental optimal scheme for measuring spectral distortions, both with and without thermal foregrounds. No exotic measurement scheme exploiting quantum resources like entanglement is required or helpful to optimally extract spectral distortions from a CMB measurement. 

The optimality of photon counting follows simply from the structure of the density matrix for the CMB photons, namely that it is diagonal in the photon number basis. This in particular means that there is no phase coherence between different photon number states. It would be interesting to consider possible cosmological signatures with coherent phase information, in which case a more sophisticated quantum measurement strategy may be advantageous.

The natural remaining question is how close one could actually get to implementing such an idealized measurement. As case studies, we include a few comments comparing the ideal case to the proposed PIXIE and SPECTER missions in Appendix ~\ref{PIXIE_SPECTER}. In principle, the binned ideal photon counting strategy requires two key ingredients: frequency binning with independent measurements, and no photodetector noise. Collecting microwave photons and routing them into frequency bins should not present a fundamental roadblock, and is currently being developed for many purposes, notably line intensity mapping measurements~\cite{robson2022simulation,becerril2026sumac}. The more difficult issues lie with low-energy photodetection. One is how to calibrate the instrument so that the expected photon detection signal in each band is precisely known. The other is how to achieve ``noiseless'' photodetection at CMB frequencies, where thermal detector fluctuations lead to spurious dark count events. This is the subject of intensive work in many detector architectures~\cite{chen2011microwave,esmaeil2021superconducting,haldar2024continuous,smith2024mkidgen3,fink2024superconducting}, and we hope that our results here provide further motivation for this.

\section*{Acknowledgments}
We thank Ritoban Basu Thakur, Colin Hill, Johannes Hubmayr, Joel Meyers, Alina Sabyr, and Aritoki Suzuki for helpful discussions. Claude Opus 5.5 was used for proofreading the paper and basic search tasks. Our work at Berkeley Lab is supported by the U.S. Department of Energy (DOE), Office of High Energy Physics, under Contract No. DEAC02-05CH11231, by DOE Quantum Information Science Enabled Discovery (QuantISED) for High Energy Physics, and by the Heising-Simons Foundation under grant 2023-4467 ``Testing the Quantum Coherence of Gravity''.

\bibliography{references.bib}
\bibliographystyle{utphys.bst}

\clearpage
\appendix
\section{$y-$distortions}
\label{y-distortions}
The mean number of photons in a single mode of the radiation field due to the $y-$distortions is
\begin{align}
    \begin{split}
        & \Delta n_y(x,y,\theta_e) = y\bigg(\frac{xe^{x}}{(e^{x}-1)^2}\bigg(x\coth\bigg(\frac{x}{2}\bigg)-4\bigg) + Y_1(x) \theta_{e} \\& + Y_2(x)\theta_e^2(1+\omega_2^{\rm{eSZ}}) + Y_3(x) \theta_e^3(1+3\omega_2^{\rm{eSZ}})\bigg).
        \label{y-distortions-rel-corrections}
    \end{split}
\end{align}
For a given mode of the radiation field, $x=\nu_{\mb{k}}/T_{\rm{CMB}}$.  $\theta_e = {T_{\rm{eSZ}}}/{m_e}$ characterizes the strength of the relativistic corrections to the $y-$distortions. The $Y_{i}(x)$ functions are obtained from the asymptotic expansions of the relativistic corrections to the $y-$distortion signal \cite{sazonov1998cosmic,itoh1998relativistic,challinor1998relativistic,abitbol2017prospects}.  $\omega_2^{\rm{eSZ}}$ is a relativistic correction parameter, which is not treated as a free parameter for the Fisher forecast.
\section{Foreground models}
\label{foreground_models}
The presence of astrophysical foregrounds is a major challenge for measuring spectral distortions and significantly increases the required observation time. As shown in Fig.~\ref{fig:CMB_signal}, the difference between the CMB signal when $\mu=2\times10^{-8}$ (the $\Lambda$CDM value) and $\mu = 0$ is several orders of magnitude smaller than these foregrounds. For our numerical estimates, we consider foreground models with the fiducial parameters used in ~\cite{abitbol2017prospects}, summarized in Table~\ref{tab:foregrounds}. 

The noise in each frequency band depends on the total number of photons, $n_{\rm{tot}}(\nu_{\mb{k}})$, incident on the detector in a given frequency band in a given observation time from the CMB and foregrounds. The noise includes a bunching term in addition to the photon shot noise. We assume that the noise is uncorrelated across distinct frequency bands and proceed to compute the classical Fisher information for frequency binned intensity measurements as in the foreground-free case.

\begin{table*}[t]
\centering
\small
\begin{tabular}{|c|c|c|}
\hline
Foreground & Spectral Radiance [Jy/sr] &
Free Parameters and Fiducial Values  \\
\hline

Thermal Dust &
$\displaystyle
\begin{gathered}
\Delta I_{\mathrm D}(\nu)
=A_{\mathrm D}x^{\beta_{\mathrm D}}
\frac{x^3}{e^x-1}; \hspace{10 pt} x=\frac{h\nu}{kT_{\mathrm D}}
\end{gathered}
$ &
$\displaystyle
\begin{gathered}
A_{\mathrm D}=1.36\times10^6~\mathrm{Jy/sr}\\
\beta_{\mathrm D}=1.53\\
T_{\mathrm D}=21~\mathrm K
\end{gathered}
$ 

\\[12pt]
\hline

CIB &
$\displaystyle
\begin{gathered}
\Delta I_{\mathrm{CIB}}(\nu)
=A_{\mathrm{CIB}}x^{\beta_{\mathrm{CIB}}}
\frac{x^3}{e^x-1};\hspace{10 pt} x=\frac{h\nu}{kT_{\mathrm D}}
\end{gathered}
$ &
$\displaystyle
\begin{gathered}
A_{\mathrm{CIB}}=3.46\times10^5~\mathrm{Jy/sr}\\
\beta_{\mathrm{CIB}}=0.86\\
T_{\mathrm{CIB}}=18.8~\mathrm K
\end{gathered}
$ 
\\[12pt]
\hline

Synchrotron &
$\displaystyle
\Delta I_{\mathrm S}(\nu)
=A_{\mathrm S}\left(\frac{\nu}{\nu_0}\right)^{\alpha_{\mathrm S}}
\left[1+\frac12\omega_{\mathrm S}
\ln^2\left(\frac{\nu}{\nu_0}\right)\right]
$ &
$\displaystyle
\begin{gathered}
A_{\mathrm S}=288.0~\mathrm{Jy/sr}\\
\alpha_{\mathrm S}=-0.82\\
\omega_{\mathrm S}=0.2
\end{gathered}
$ 

\\[12pt]
\hline

Free-Free &
$\displaystyle
\begin{gathered}
\Delta I_{\mathrm{FF}}(\nu)
=A_{\mathrm{FF}}\left(
1+\ln\left[1+
\left(\frac{\nu_{\mathrm{ff}}}{\nu}\right)^{\sqrt{3}/{\pi}}
\right]\right)
\end{gathered}
$ &
$\displaystyle
\begin{gathered}
A_{\mathrm{FF}}=300~\mathrm{Jy/sr}\\
\end{gathered}$
\\[12pt]
\hline
Integrated CO & $\Delta I_{\rm{CO}} = A_{\rm{CO}}\times $ CO template $(\nu)$ & $A_{\rm{CO}}=1$\\
\hline
Spinning dust & $\Delta I_{\rm{SD}} = A_{\rm{SD}}\times $ SD template $(\nu)$ & $A_{\rm{CO}}=1$ \\
\hline
\end{tabular}
\caption{Foreground spectral radiance models and their parameters \cite{abitbol2017prospects}. The fixed reference frequencies for the synchrotron foreground and the free-free emission foreground are $\nu_0 = 100~\rm{GHz}$ and $\nu_{\rm{ff}}=4728.78 ~\rm{GHz}$. The CO templates and SD templates were taken from \cite{sd_foregrounds}.}
\label{tab:foregrounds}
\end{table*}

\section{Fourier Transform Spectroscopy}
\label{fts_section}
Fourier transform spectroscopy measures the spectral decomposition of the intensity of a source of radiation using the Fourier transform of an interferogram. Instead of measuring the intensity in independent frequency channels, the intensity at the output port of an interferometer is measured for varying optical path differences, $x_n$, between the arms of the interferometer. For the nulling Martin-Puplett FTS proposed in the PIXIE concept, the sum of the intensities measured at the two left detectors $I_L(x_n)$ and the two right detectors $I_R(x_n)$ is \cite{kogut2011primordial}
\begin{eqnarray}
\label{intensity_fts}
    I_{L(R)}(x_n) & = &\frac{1}{2}\sum_k \Delta \nu [I_{\rm{SKY}}(\nu_\textbf{k},\boldsymbol{\theta})+ I_{\rm{BBR}}(\nu_\textbf{k}) \nonumber \\
    & \pm & (I_{\rm{SKY}}(\nu_\textbf{k},\boldsymbol{\theta})-I_{\rm{BBR}}(\nu_\textbf{k}))
    \cos(2\pi\nu_\textbf{k} x_n)].\nonumber \\
\end{eqnarray}
Here, $I_{\rm{SKY}}(\nu_\textbf{k}, \boldsymbol{\theta})$ is the incident spectral intensity density from the CMB and foregrounds and $I_{\rm{BBR}}(\nu_\textbf{k})$ is the spectral intensity density of a reference load, assumed to be a perfect blackbody at the CMB temperature. A more detailed approach to designing the reference load, accounting for deviations from a blackbody spectrum, can also be considered \cite{aghanim2026fossilfuturemissioncmb}.

The uncertainty in the intensity $\sigma^2_{I_{L(R)}(x_n)}$ has contributions from the CMB intensity, the foregrounds, and the reference blackbody over the entire spectrum,
\begin{equation}
\label{fts_noise}
 \begin{split}
     &\sigma^2_{I_{L(R)}(x_n)} = \sum_{\nu_\textbf{k}} 2 \sigma^2_{n_{\rm{tot}(\nu_\textbf{k})}}\bigg(\frac{h\nu_\textbf{k}}{A\Omega \Delta \nu T_{\rm{obs}}}\bigg)^2.\\
 \end{split}
\end{equation}
Here $\sigma^2_{n_{\rm{tot}(\nu_\textbf{k})}}$ is the noise due to all incident photons in the band.
We sweep over optical path differences $x_n$, and add the Fisher information from each of these measurements. For a given $x_n$, the Fisher information matrix is
\begin{equation}
    F^{ij}(x_n,\boldsymbol{\theta}) = \sum_{k=L,R} \frac{1}{\sigma_{I_k(x_n)}^2}\bigg(\frac{\partial I_k(x_n)}{\partial \theta_i}\bigg)\bigg(\frac{\partial I_j(x_n)}{\partial \theta_j}\bigg).
\end{equation}

From the total Fisher information matrix, we can compute $T_{\rm{obs}}$ as done in Eq.~\eqref{T_obs_ineq}. The total integration time is $T_{\rm{tot}} = N \times T_{\rm{obs}}$ where $N$ is the number of path differences sampled.

\section{Comparison with mission concepts}
We compare proposed mission concepts to idealized photon counting and FTS detectors. For photon counting, we consider 20 broadband feedhorns that span the entire frequency range. The radiation incident on each broadband feedhorn gets further divided into narrower bins for finer frequency resolution. While Fig.~\ref{fig:missions_compare} assumes full sky measurements, here we consider the observed fraction of the sky $f_{\rm{sky}}=0.7$ to compare to~\cite{kogut2025primordial,abitbol2017prospects,sabyr2025specter}. 
\begin{figure*}[htbp]
    \centering
    \begin{minipage}[b]{0.44\textwidth}
        \centering
        \includegraphics[width=\linewidth]{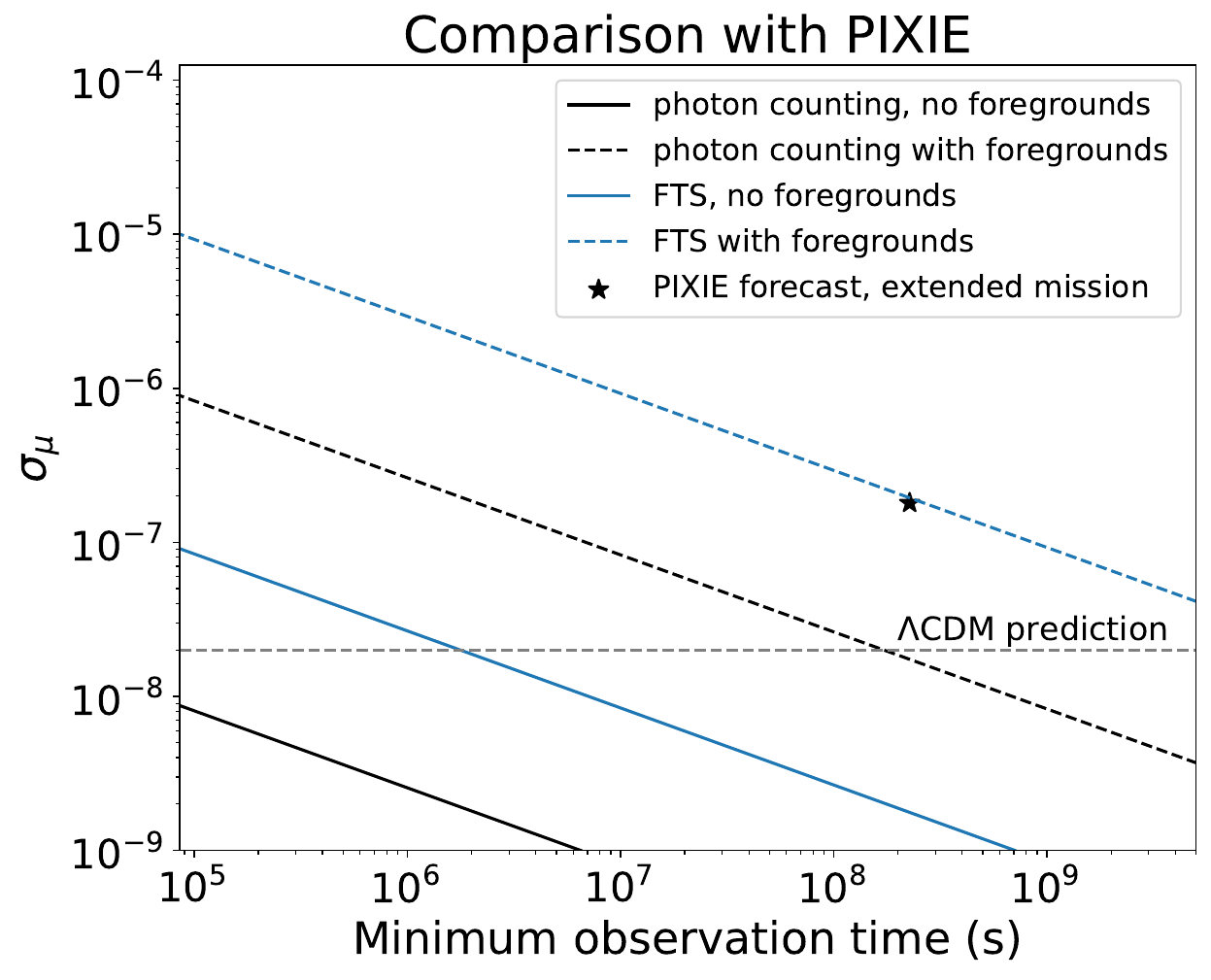}
    \end{minipage}
    \hfill 
    \begin{minipage}[b]{0.44\textwidth}
        \centering
        \includegraphics[width=\linewidth]{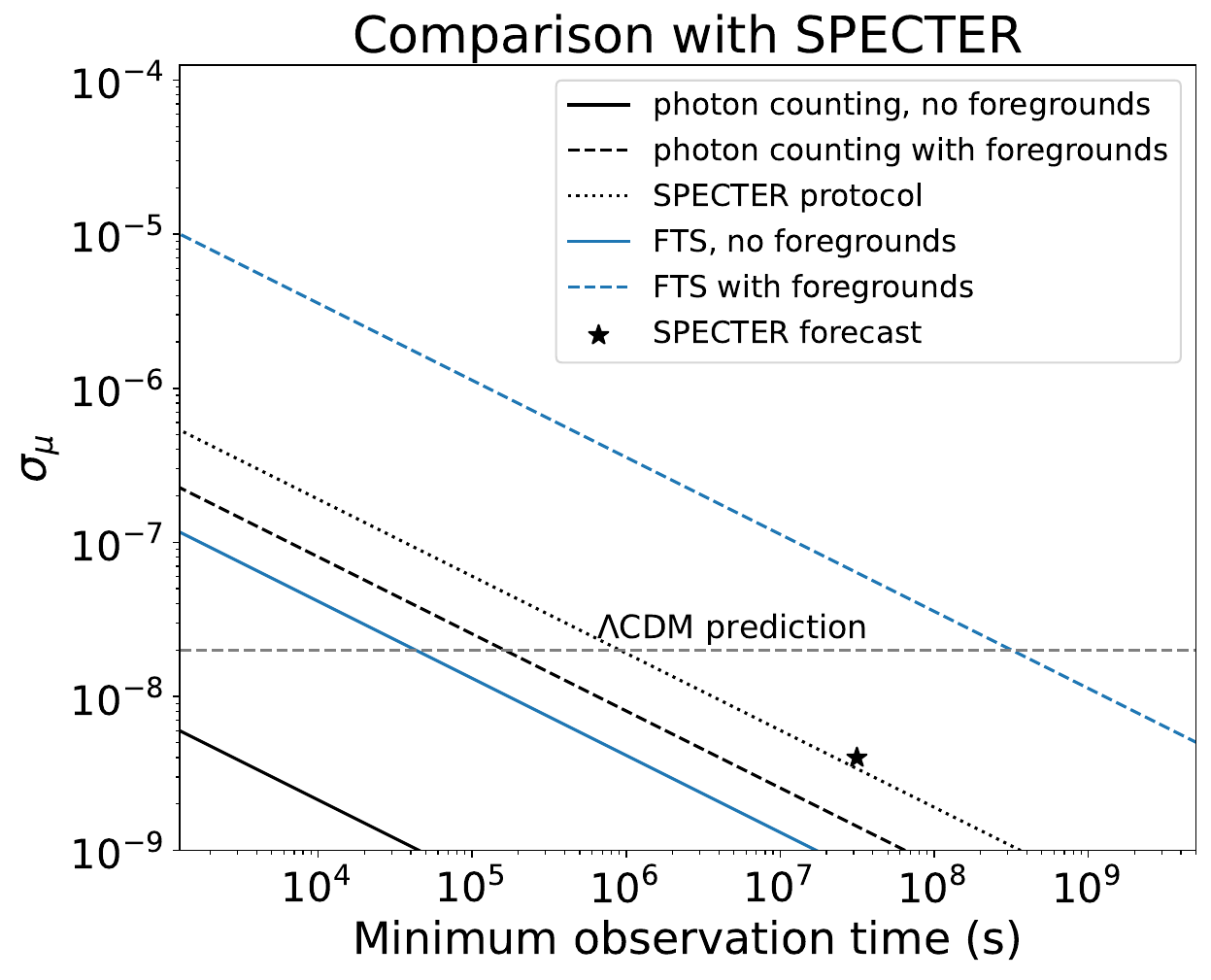}
    \end{minipage}
\caption{Comparison of the obtainable measurement uncertainty $\sigma_{\mu}$ in the $\mu$-distortion as a function of observation time with the forecasts from mission concepts PIXIE (left) and SPECTER (right). The various lines represent idealized photon-counting and Fourier Transform Spectrometer (FTS) detectors, using foreground parameters in Table~\ref{tab:foregrounds}. For the comparison with PIXIE the frequency range $15~\rm{ GHz} - 3 ~\rm{THz}$, bandwidth $15 ~\rm{GHz}$, and total \'etendue $16 ~\rm{cm^2 sr}$ are taken from~\cite{abitbol2017prospects}. For the comparison with SPECTER, the frequency range $1  ~\rm{ GHz} - 2 ~\rm{ THz}$ is taken from the SPECTER mission concept~\cite{sabyr2025specter}. The PIXIE and SPECTER markers represent the level up to which the uncertainty in $\mu$ can be constrained for the extended PIXIE mission and SPECTER given the time for which the spectral distortions are to be observed. The SPECTER protocol line corresponds to photon counting with 16 frequency bands with frequency dependent \'etendue, described in~\cite{sabyr2025specter}. }
\label{fig:PIXIE_SPECTER_comparison}
\end{figure*}
\label{PIXIE_SPECTER}

\textbf{PIXIE:} This mission concept proposes to use Fourier transform spectroscopy (FTS) \cite{kogut2011primordial} which increases the necessary integration time compared to frequency resolved intensity measurements (see Fig.~\ref{fig:PIXIE_SPECTER_comparison}). For direct frequency resolved intensity measurements, the noise in any channel is only due to the number of photons incident in that channel, whereas for the FTS all the incident photons across the entire frequency range contribute to the noise in the signal as calculated in Eq. ~\eqref{fts_noise}. This is relevant when the bolometer noise, including dark counts, is predicted to be below the photon noise ~\cite{nagler2016multimode} and the photon noise becomes the limiting factor. We consider no priors on the free parameters, unlike the PIXIE forecast (starred point), which considers $10\%$ priors on  $A_S$ and $\alpha_S$ resulting in a slightly more optimistic forecast than our FTS with foregrounds (blue dashed line).

\textbf{SPECTER:} This mission concept proposes to measure the CMB spectral distortions~\cite{sabyr2025specter} using binned intensity measurements, optimized over the bandwidths and the number of detectors in each frequency band to minimize the total detector area. The total \'etendue,  $1976 ~\rm{cm}^2 \rm{sr}$, for the photon counting and FTS lines is taken to match the \'etendue across all the feedhorns in SPECTER. The photon counting protocol considers 400 frequency bands with a uniform $5~\rm{GHz}$ bandwidth  in contrast to SPECTER which has 16 channels with varying bandwidths. Moreover, for the photon counting lines, the total \'etendue is spread equally across all feedhorns, yielding an etendue of $\sim 100 ~\rm{cm^2 sr}$ per feedhorn, as opposed to SPECTER, where the \'etendue varies across different frequencies, which also affects the Fisher forecasts. Unlike the SPECTER forecast, our calculations do not consider the thermal bolometric or electronic readout noise, which accounts for the small discrepancy between their predicted sensitivity (starred point) and the dotted line.
\end{document}